\documentclass[a4paper,11pt]{article}
\pdfoutput=1

\usepackage{jheppub}
\usepackage[T1]{fontenc}

\usepackage{booktabs}
\usepackage{array}
\usepackage{bm}
\usepackage{colortbl}
\usepackage{float}
\usepackage{mathtools}
\usepackage{multirow}
\usepackage{siunitx}
\usepackage{tabularx}

\DeclareSymbolFont{usualmathcal}{OMS}{cmsy}{m}{n}
\DeclareSymbolFontAlphabet{\mathcal}{usualmathcal}

\numberwithin{equation}{section}

\newlength\savewidth
\newcommand\shline{\noalign{\global\savewidth\arrayrulewidth
  \global\arrayrulewidth 1pt}\hline\noalign{\global\arrayrulewidth\savewidth}}
\newcommand\midline{\noalign{\global\savewidth\arrayrulewidth
  \global\arrayrulewidth 0.5pt}\hline\noalign{\global\arrayrulewidth\savewidth}}
\newcommand{\tablestyle}[2]{\setlength{\tabcolsep}{#1}\renewcommand{\arraystretch}{#2}\centering\footnotesize}

\newcolumntype{x}[1]{>{\centering\arraybackslash}p{#1pt}}
\newcolumntype{y}[1]{>{\raggedright\arraybackslash}p{#1pt}}
\newcolumntype{z}[1]{>{\raggedleft\arraybackslash}p{#1pt}}

\newcommand{\x}{\bm{x}}
\newcommand{\e}{\bm{\epsilon}}
\newcommand{\z}{\bm{z}}
\newcommand{\vv}{\bm{v}}

\title{\boldmath CaloArt: Large-Patch \(\x\)-Prediction Transformers for High-Granularity Calorimeter Shower Generation}

\author[a,b]{Zhengkun Huang}
\author[a,b]{Gongxing Sun}

\affiliation[a]{Institute of High Energy Physics, Chinese Academy of Sciences, China}
\affiliation[b]{University of Chinese Academy of Sciences, China}

\emailAdd{huangzk@ihep.ac.cn}
\emailAdd{sungx@ihep.ac.cn}

\abstract{High-granularity calorimeters make ML-based fast shower simulation a high-dimensional
generative modeling problem, where voxel-space generators must balance
physics fidelity with training and inference cost. This work studies
large-patch tokenization with \(\x\)-prediction, enabling efficient raw voxel
generation. 
We propose CaloArt, a modernized DiT-style backbone with 3D positional encoding and architectural refinements, 
trained via conditional flow matching with decoupled prediction and loss spaces.
On CaloChallenge Dataset 2, where small patch size remains affordable, \(\vv\)-prediction performs well, 
and CaloArt achieves the best FPD, strongest high-level and ResNet classifier
metrics. On CaloChallenge Dataset 3, the
\(40500\)-voxel grid makes large patches necessary; \(\x\)-prediction improves
all reported metrics over \(\vv\)-prediction and places CaloArt on the
quality-generation-time Pareto frontier. The final CCD2 and CCD3 models both retain
\(\mathcal{O}(10)\,\mathrm{ms}\) single-GPU generation time, with \(9.71\) and
\(11.14\,\mathrm{ms}\) per shower.
These results support large-patch voxel-space diffusion transformers
with \(\x\)-prediction as a compute-efficient route to high-granularity
calorimeter shower synthesis, reducing training and inference cost without
a pretrained latent tokenizer.
}

\begin{document}
\setcounter{tocdepth}{2}
\maketitle
\clearpage
\flushbottom

\section{Introduction}
\label{sec:intro}

Detector simulation is essential for interpreting collider data and comparing
experimental measurements with theoretical predictions. Geant4-based calorimeter shower simulation remains a major computational
bottleneck in large scale Monte Carlo production and is a
particularly important target for acceleration~\cite{2003_Agostinelli_NIMA_Geant4,2016_Allison_NIMA_Geant4}.
This pressure will increase at the high-luminosity Large Hadron Collider
(LHC) and future colliders, where substantially larger simulated event samples
will be required. In parallel, modern calorimeter designs are moving toward finer readout
granularity, increasing the dimensionality of shower representations~\cite{2022_Wiehe_NIMA_CMS_HGCAL,2024_ALICE_FoCal_TDR,2016_Sefkow_RMP_CALICE,2020_ILD_arXiv,2021_Aleksa_EPJPlus_FCCee}.
Fast calorimeter generators must therefore be accurate and efficient while
scaling to high-dimensional shower data.

ML-based fast simulation methods can be broadly divided, at the level of shower
representation, into grid voxel and point cloud approaches. The point cloud representation has been motivated
as a natural way to describe sparse showers in
highly granular calorimeters. In this formulation, point cloud models generate energy-deposition points and
are less directly tied to the total number of readout
cells~\cite{2023_Buhmann_JINST_CaloClouds,2024_Buhmann_JINST_CaloCloudsII,2026_Buss_JINST_CaloClouds3,2025_Birk_arXiv_2501_05534}.
Voxel-space generation is a complementary readout-aligned setting, directly
modeling per cell energy deposits used in grid voxel benchmarks such as
the CaloChallenge~\cite{2025_Krause_RPP_88}. Voxel-space models, whether based
on convolutional U-Net
backbones~\cite{2015_Ronneberger_MICCAI_UNet,2022_Mikuni_PRD_CaloScore,2024_Mikuni_JINST_19_P02001,2023_Amram_PRD_CaloDiffusion}
or transformer/ViT backbones~\cite{Favaro:2024rle,Favaro:2026awn}, face
rapidly increasing computational cost as the voxel count grows, as reflected
in the CaloChallenge Dataset 2 (CCD2) -- CaloChallenge Dataset 3 (CCD3) timing
comparison~\cite{2025_Krause_RPP_88}. Latent-space voxel generators have
therefore been
explored~\cite{2023_Madula_CaloLatent,2024_Liu_arXiv_CaloVQ,Favaro:2024rle},
but calorimeter showers lack a standard perceptual representation analogous to
VGG~\cite{2014_Simonyan_arXiv_VGG} or
DINOv2~\cite{2023_Oquab_arXiv_DINOv2} image features, making a high-fidelity
tokenizer difficult to train.

This work therefore focuses on direct voxel-space generation without introducing an autoencoder tokenizer. 
Instead, we control the backbone computational cost through large patch sizes, reducing the
token count and forward Gflops in high-resolution shower generation. To make
such large-patch models train reliably, we adopt a
\(\x\)-prediction formulation that directly predicts the clean sample. 
This combination is particularly important for CCD3, where the much larger voxel grid makes large
patch sizes necessary under realistic compute budgets. 

The use of \(\x\)-prediction is motivated by the JiT~\cite{2025_Li_arXiv_2511_13720} perspective on
diffusion transformers~\cite{2022_Peebles_arXiv_2212_09748} in high-dimensional pixel spaces. Modern diffusion models commonly predict
either the noise (\(\e\)-prediction)~\cite{2020_Ho_NeurIPS_DDPM,2021_Song_ICLR_DDIM,2021_Song_ICLR},
or the flow velocity (\(\vv\)-prediction)~\cite{2022_Salimans_ICLR_ProgressiveDistillation,2023_Lipman_ICLR,
2023_Liu_RectifiedFlow,2023_Albergo_ICLR}, a quantity that combines clean data
and noise. Under the manifold assumption, clean samples are expected to lie
on a low-dimensional data manifold, whereas noise and noised quantities do
not~\cite{2006_Chapelle_Book,2009_Carlsson_BAMS,2010_Vincent_JMLR}.
This distinction becomes especially important for large-patch diffusion transformers,
where the raw patch dimensionality can approach or exceed the model width,
making direct prediction of clean data a natural choice, since clean data are
assumed to lie on a manifold
~\cite{2025_Li_arXiv_2511_13720}.

To align the large-patch raw pixel image generation perspective with 3D  
voxelized calorimeter showers, we propose CaloArt, a modernized DiT-style  
backbone for direct raw voxel calorimeter generation. Our formulation  
explicitly separates the prediction space from the loss space, allowing us to  
compare \(\x\)-prediction and \(\vv\)-prediction under matched calorimeter generation  
settings on CCD2 and CCD3. CCD2 contains \(6480\) voxels, a scale close to common image latent-diffusion  
dimensions~\cite{2022_Rombach_CVPR_LDM,2022_Peebles_arXiv_2212_09748}.
Its smaller voxel count leaves sufficient computational budget for  
relatively finer 3D patch tokenization, where \(\vv\)-prediction remains the  
stronger choice. CCD3 is qualitatively different. Its \(40500\)-voxel grid places direct
full shower generation closer to raw pixel image generation
~\cite{2021_Dhariwal_NeurIPS_ADM,2024_Crowson_arXiv_HDiT,
2025_Yu_arXiv_PixelDiT}, making large 3D patches necessary under practical
training and inference budgets. Under this constraint, \(\x\)-prediction makes such large patches trainable, keeps the
backbone Gflops small, and improves over \(\vv\)-prediction across all reported
metrics. Since a tokenizer for calorimeter showers is difficult to design,
large-patch tokenization with \(\x\)-prediction provides an alternative route
toward ``democratizing'' high-granularity calorimeter shower
synthesis~\cite{2022_Rombach_CVPR_LDM}: direct raw-voxel generation without a
learned autoencoder tokenizer. In our final CaloArt CCD3 result, the reported
model is trained on a single NVIDIA A800 GPU in \(17.57\) hours.

This paper is organized as follows. Section~\ref{sec:method} presents
conditional flow matching with decoupled prediction and loss spaces, together with the CaloArt backbone. Section~\ref{sec:Gflops} defines the quality and
compute metrics. Sections~\ref{sec:ccd2_results} and~\ref{sec:ccd3_results}
report the CCD2 and CCD3 experiments and results, including the Gflops scaling,
prediction-target, and backbone ablations. Section~\ref{sec:Conclusion}
concludes the paper, and the appendices provide supporting plots,
derivations, and hyperparameters.

\section{Method}
\label{sec:method}

\subsection{Conditional flow matching and prediction targets}

Conditional flow matching (CFM)~\cite{2023_Lipman_ICLR} is used as the generative formulation throughout this work, 
with $\vv$-prediction on CCD2 and $\x$-prediction on CCD3. 
The choice of prediction target is governed by the trade-offs among generation quality, training cost, and sampling efficiency. 
More generally, diffusion models can be formulated in the space of \(\x\), \(\e\), or \(\vv\). 
The choice of space determines not only where the loss is defined (\textbf{loss space}), but also what the network predicts (\textbf{prediction space}). 
Once these two choices are decoupled, multiple valid pairings arise, 
and different pairings can lead to markedly different optimization behavior and generation quality in practice.

To make the derivations and the transformations between \(\x\), \(\e\), and \(\vv\) explicit, 
we adopt the standard linear interpolation path between data and noise, 
following common flow-based formulations~\cite{2023_Lipman_ICLR,2023_Albergo_ICLR,2023_Liu_RectifiedFlow}. Let \(\x \sim p_{\mathrm{data}}(\x \mid c)\) denote a shower tensor conditioned on physical variables \(c\), and let \(\e \sim \mathcal{N}(0,I)\) be a Gaussian noise sample of the same shape. During training, a noisy sample \(\z_t\) is an interpolation of the form \(\z_t = a_t \x + b_t \e\), where \(a_t\) and \(b_t\) are predefined noise schedules at time \(t \in [0,1]\). In this work, we use the linear schedule \(a_t=t\) and \(b_t=1-t\), which gives
\begin{equation}
\z_t = t \x + (1-t)\e, \qquad t \in [0,1].
\label{eq:linear-path}
\end{equation}
This leads to \(\z_t \sim p_{\mathrm{data}}\) when \(t=1\). The time variable \(t\) is sampled from a logit-normal distribution, 
with \(\mathrm{logit}(t) \sim \mathcal{N}(\mu,\sigma^2)\)~\cite{2024_Esser_arXiv_2403_03206}. 
A flow velocity \(\vv\) is defined as the time derivative of \(\z_t\), that is, \(\vv = \z'_t = a_t' \x + b_t' \e\). Under the linear schedule above, this gives
\begin{equation}
\vv = \x - \e.
\label{eq:linear-velocity}
\end{equation}
The flow-based formulation minimizes the loss
\begin{equation}
L = \mathbb{E}_{t,\x,\e}\left\| \vv_\theta(\z_t,t,c) - \vv \right\|^2,
\label{eq:v-loss}
\end{equation}
where \(\vv_\theta\) is a function parameterized by \(\theta\). Given the learned velocity field \(\vv_\theta\), sampling is performed by solving an ordinary differential equation (ODE) for \(\z\)~\cite{2023_Lipman_ICLR,2023_Albergo_ICLR,2023_Liu_RectifiedFlow},
\begin{equation}
d\z_t / dt = \vv_\theta(\z_t,t,c),
\label{eq:sampling-ode}
\end{equation}
starting from \(\z_0 \sim \mathcal{N}(0,I)\) and ending at \(t=1\). In practice, this ODE is approximated with numerical solvers. By default, we use a 32-step Heun solver.

\paragraph*{\textbf{Prediction targets and loss space.}}

The network's direct output can be defined in the \(\x\), \(\e\), or \(\vv\)-space. 
Given Eqs.~\eqref{eq:linear-path} and~\eqref{eq:linear-velocity}, once one of \(\{\x,\e,\vv\}\) is predicted, the other two can be inferred deterministically. 
Likewise, the training objective can be defined in any of these spaces, and the prediction space and loss space need not coincide. 
For completeness, Appendix~\ref{app:xev} summarizes all prediction-space / loss-space combinations under the linear path above, together with the corresponding transformations. 
In particular, Eq.~\eqref{eq:v-loss} corresponds to the standard $\vv$-prediction with $\vv$-loss formulation.

The additional case considered in this work is $\x$-prediction with $\vv$-loss, where the network directly predicts the clean sample
\begin{equation}
\x_\theta = \mathrm{net}_\theta(\z_t,t,c).
\label{eq:x-prediction}
\end{equation}
Using Eqs.~\eqref{eq:linear-path} and~\eqref{eq:linear-velocity}, this output is mapped to the corresponding velocity prediction,
\begin{equation}
\vv_\theta = (\x_\theta-\z_t)/(1-t).
\label{eq:x-to-v-prediction}
\end{equation}
Although the network output is defined in the \(\x\)-space, the loss is still evaluated in the \(\vv\)-space. Using the relation in Eq.~\eqref{eq:x-to-v-prediction}, together with the target \(\vv = (\x-\z_t)/(1-t)\), the \(\vv\)-loss becomes
\begin{equation}
L = \mathbb{E}_{t,\x,\e}\left\| \vv_\theta(\z_t,t,c) - \vv \right\|^2
  = \mathbb{E}_{t,\x,\e}\frac{1}{(1-t)^2}\left\| \x_\theta(\z_t,t,c) - \x \right\|^2.
\label{eq:xpred-vloss}
\end{equation}
Hence, this $\vv$-loss can be viewed as a reweighted form of the $\x$-loss. 
At inference time, generation still proceeds in the \(\vv\)-space: the \(\x\)-space network output \(\x_\theta\) is first transformed to \(\vv_\theta\), and the same ODE in Eq.~\eqref{eq:sampling-ode} is then solved.

\begin{figure}[t]
\centering
\makebox[\linewidth][c]{\hspace*{3em}\includegraphics[width=0.85\linewidth]{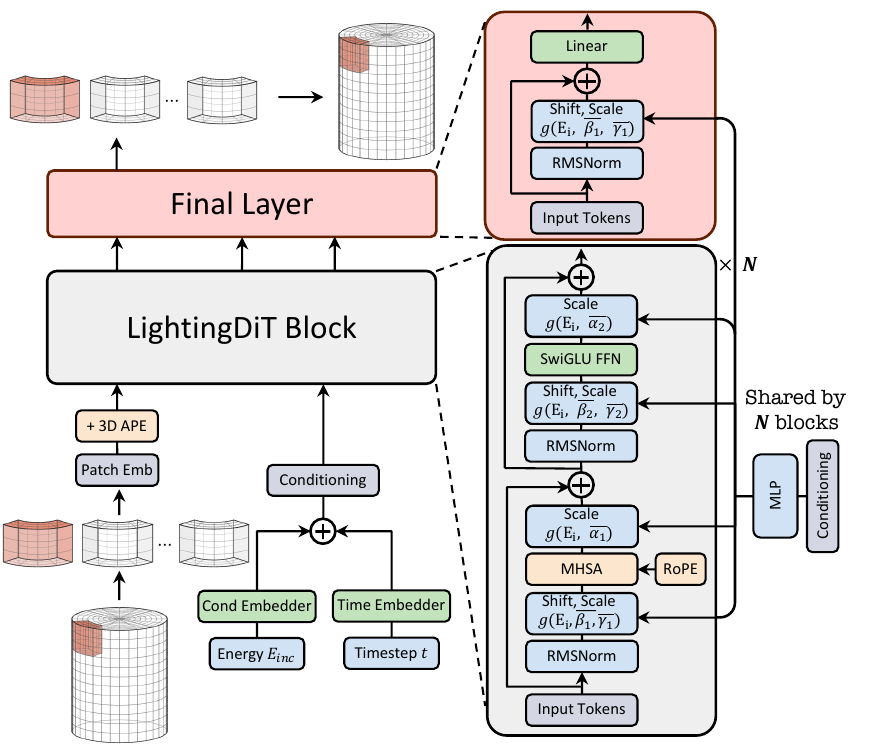}}
\caption{\textbf{Overview of the CaloArt backbone.} To improve parameter efficiency, 
all transformer blocks share the same adaLN-single modulation branch. 
The function $g(E_i,\bar{\beta},\bar{\gamma})$ denotes the block-specific combination of the shared shift and scale with the trainable embedding $E_i$ of the $i$-th transformer block. 
The resulting shift and scale parameters are used by adaptive normalization in that block.}
\label{fig:caloart_backbone}
\end{figure}

\subsection{CaloArt backbone}

CaloArt adopts a DiT-style transformer as the base architecture for direct 3D voxel-shower generation. 
Unlike latent-diffusion approaches, it operates directly on calorimeter voxels and does not introduce a separate autoencoder tokenizer. 
The input shower is first divided into non-overlapping 3D patches, embedded into a sequence of tokens, processed by a stack of transformer blocks, and finally mapped back to the original shower geometry through an unpatchify operation. The model is conditioned directly on incident-particle information, such as \(E_{\mathrm{inc}}\), without requiring additional derived conditioning variables such as \(E_{\mathrm{layer}}^i\). 
Compared with previous transformer backbones for calorimeter simulation~\cite{2025_Raikwar_arXiv_2509_07700,Favaro:2025ift,Favaro:2026awn}, CaloArt introduces the following practical architectural refinements:

\paragraph*{\textbf{Shared conditioning modulation.}} The physical conditioning variables are embedded into a model-width vector, which is then added to the timestep embedding to form a global conditioning representation. 
This conditioning signal is injected into each transformer block through \textit{adaLN-Zero} modulation. 
In addition to the standard DiT design, we consider a PixArt-style shared modulation design~\cite{2024_Chen_PixArtAlpha}, related to the \textit{adaLN-single} strategy. 
Instead of using a separate modulation projection in every block, 
a shared modulation projection first computes a tuple of all the scale and shift parameters in \textit{adaLN}, 
which is then combined with layer-specific trainable embeddings to produce the block-specific modulation parameters. 
In the ablation studied in Section~\ref{subsec:shared_modulation_ablation}, this reduces the parameter count from \(26.13\)M to \(18.77\)M while keeping the main generation metrics close to the non-shared baseline.

\paragraph*{\textbf{3D positional encoding.}} To encode the spatial structure of the shower, CaloArt uses a 3D axial rotary positional embedding. 
Each patch token is assigned a coordinate \((z,r,\alpha)\), and RoPE phases are constructed separately along the three axes using the standard RoPE frequency schedule~\cite{2021_Su_RoFormer}. 
The rotary dimensions are divided across the three axes and applied to the query and key features in the attention module, making attention interactions explicitly dependent on relative 3D patch positions. 
The backbone supports three positional configurations: \texttt{ape-only}, \texttt{rope-only}, and \texttt{ape+rope}. 
The absolute component is implemented as 3D sinusoidal absolute positional embeddings added after patchification. 
Based on the ablation in Section~\ref{subsec:RoPE_ablation}, we use the \texttt{ape+rope} configuration as the default setting in the experiments below.

\paragraph*{\textbf{Modern transformer refinements.}} CaloArt incorporates several backbone refinements commonly used in recent task-agnostic diffusion transformers. 
Following the LightningDiT modernization recipe~\cite{2025_LightningDiT}, 
we replace the standard FFN with a SwiGLU FFN~\cite{2020_Shazeer_GLU}, replace LayerNorm with RMSNorm~\cite{2019_Zhang_RMSNorm}, and apply query--key normalization in the attention module~\cite{2020_Henry_QKNorm}.

\medskip 
Overall, CaloArt keeps the direct full-shower generation setup of CaloDiT-style models, but strengthens the backbone through practical updates to positional encoding, conditioning modulation, and transformer block components. The resulting model is intended as a stronger default DiT backbone for voxel-based calorimeter generation.

\subsection{Preprocessing}

For shower voxels, preprocessing follows CaloDiT-v2~\cite{2025_Raikwar_arXiv_2509_07700}. 
Voxel energies below \(15.15\,\mathrm{keV}\) are set to zero, and the remaining values are transformed by a logarithm followed by global standardization:
\begin{equation}
\tilde{x}_i=\frac{\log(x_i+\varepsilon)-\mu}{\sigma},
\end{equation}
where \(\varepsilon=10^{-6}\), and \(\mu\) and \(\sigma\) are scalar normalization constants shared across all voxels. 
Conditioning is based only on the incident energy \(E_{\mathrm{inc}}\), which is mapped by a logarithmic transform followed by scaling to \([0,1]\)~\cite{2021_Krause_arXiv_2106_05285,Favaro:2024rle,2024_Mikuni_JINST_19_P02001}.

For CaloChallenge Dataset 3, generation occasionally produced outliers with unphysically large deposited energies. 
Such cases can arise when large voxel values appear near the end of ODE sampling and are then amplified by the inverse logarithmic transform. 
To mitigate this issue, a redraw strategy similar to that used in prior calorimeter-generation work is adopted~\cite{2024_Buss_arXiv_2405_20407}. 
Concretely, samples with a deposited energy to incident energy ratio above \(2.7\) were rejected and redrawn. 
In the generation of \(100{,}000\) showers, this criterion rejected \(110\) samples.

\section{Evaluation Metrics and Computational Cost}
\label{sec:Gflops}

\subsection{Sample quality metrics}

The metrics reported in this paper follow the standard CaloChallenge evaluation protocol~\cite{2025_Krause_RPP_88}. 
The evaluation therefore includes the standard set of high-level shower observables used in recent calorimeter-generation work, 
including energy-related quantities such as the total deposited energy fraction $E_{\mathrm{tot}}/E_{\mathrm{inc}}$, 
per-layer deposited energies $E_i$, and average longitudinal and radial shower profiles, as well as shape observables such as the energy centroid and width along detector axes, 
together with shower sparsity~\cite{2025_Krause_RPP_88,Favaro:2026awn}. 
Appendix~\ref{app:high_level_histograms} shows the corresponding histograms of the high-level features for the final CCD2 and CCD3 models.

Distance-based metrics further complement the one-dimensional histogram comparisons. 
In particular, we consider the Fr\'echet-Physics-Distance (FPD) and Kernel-Physics-Distance (KPD), 
which are defined on the same high-level features~\cite{2023_Kansal_PRD_107_076017}. 
FPD measures the Fr\'echet distance between Gaussian fits to the generated and reference feature distributions, 
while KPD provides a complementary kernel-based comparison in the same feature space~\cite{2023_Kansal_PRD_107_076017}. 
These metrics were introduced for HEP evaluation in analogy with the Fr\'echet Inception Distance and Kernel Inception Distance~\cite{2023_Kansal_PRD_107_076017,2017_Heusel_NeurIPS,2018_Binkowski_ICLR}. 
FPD and KPD implementation follows JetNet~\cite{2023_Kansal_JOSS_8_5789} with the same hyperparameter settings as in the CaloChallenge evaluation~\cite{2025_Krause_RPP_88}. 
Since KPD is known to correlate strongly with FPD in this setting, we focus on FPD in the discussion below~\cite{2025_Krause_RPP_88,Favaro:2026awn}.

Neural classifier metrics are reported to probe sample fidelity in high dimensions. 
Specifically, the evaluation includes a high-level classifier trained on high-level observables, 
a low-level classifier trained directly on voxelized showers, and a CNN-ResNet classifier trained on the same low-level inputs but with a stronger spatial inductive bias, 
making it more sensitive to spatial mismodeling patterns in generated showers~\cite{2025_Krause_RPP_88,Favaro:2026awn}. 
For all three classifiers, the reported metric is the AUC, where values closer to $0.5$ indicate that generated showers are harder to distinguish from the Geant4 reference. 
Following the CaloChallenge prescription, each classifier is trained independently $10$ times, 
and the results in the tables below are reported as the mean together with the standard deviation over these $10$ runs~\cite{2025_Krause_RPP_88}.

\subsection{Generation time}

In addition to sample quality, the practical usefulness of a fast calorimeter surrogate also depends on its generation speed at inference time. 
We therefore report measured GPU generation time for the final CaloArt models under a fixed timing protocol aligned with the CaloChallenge benchmark setup~\cite{2025_Krause_RPP_88}.

Our timing study is restricted to GPU inference and is performed on a single NVIDIA A100-PCIE-40GB GPU. 
For both the final CCD2 and CCD3 models, we generate \(100{,}000\) showers using a batch size of \(100\). 
Unless otherwise stated, sampling is carried out with the same inference configuration used in the final model evaluation, a \(32\)-step Heun solver. 
The reported timing includes the full sampling loop under this setup, including one-time startup overheads at the beginning of generation, 
such as CUDA kernel initialization, the \texttt{torch.compile} overhead, as well as the transfer of each generated batch from GPU back to CPU.

\subsection{Gflops as a backbone-compute measure}

This paper adopts the same Gflops counting convention used in DiT, in which each multiply-add is counted as one flop; 
as in DiT, the reported Gflops refer to forward pass complexity only~\cite{2022_Peebles_arXiv_2212_09748,2023_DiT_GitHub_Issue14}. 
Expressing compute on this perspective makes the backbone scale of transformer-based calorimeter simulation models directly comparable to the ViT/DiT/SiT families~\cite{2021_Dosovitskiy_ICLR,2022_Peebles_arXiv_2212_09748,2024_Ma_ECCV}, 
where ViT-B, DiT-B, and SiT-B represent broadly matched backbone settings. 
For instance, the published pre-trained CaloDiT-2 model released with the \mbox{GEANT4/Par04} workflow uses a transformer backbone with embedding dimension $384$, $6$ DiT blocks, 
and a token sequence length of $360$, corresponding to a forward complexity of $4.43$ Gflops~\cite{2025_Raikwar_arXiv_2509_07700}. 
This is close to the scale of \mbox{DiT-B/4}, a base-scale DiT model with hidden size $768$, $12$ DiT blocks, and $64$ latent tokens, reported at $5.6$ Gflops~\cite{2022_Peebles_arXiv_2212_09748}. 
Such a comparison provides a more direct and intuitive sense of backbone scale in calorimeter simulation.

\paragraph*{\textbf{Gflops counting and validation.}}

For all models, we measure the single forward pass complexity using \texttt{DeepSpeed} profiling~\cite{2020_Rasley_KDD}. 
Because \texttt{DeepSpeed} counts one multiply-add operation as two float point operations, 
we divide the reported value by two to match the DiT-compatible convention described above. 
We further cross-check the resulting values using \texttt{ptflops} with the
\texttt{aten} backend~\cite{2024_Sovrasov_ptflops} and an independent
module-level complexity calculation. 
These checks give consistent results for the transformer backbones considered
here. 
Unless otherwise stated, the Gflops values reported below are the DiT-compatible values obtained from \texttt{DeepSpeed}; 
Appendix~\ref{app:gflops_validation} gives a representative validation example.

\paragraph*{\textbf{Dependence of total inference compute on solver steps.}}

For diffusion- and flow-based calorimeter generators, except for aggressively distilled one-step variants~\cite{2025_Raikwar_arXiv_2509_07700}, 
inference typically requires repeated backbone evaluations to numerically solve an underlying ODE or SDE~\cite{2021_Song_ICLR,2022_Karras_NeurIPS_35,2023_Lipman_ICLR}. 
The resulting generation cost therefore depends not only on the Gflops of a single backbone forward pass, 
but also on the solver step count and the number of backbone evaluations required per step.
For a solver with $S$ steps, the fourth-order Runge--Kutta (RK4) method requires $4S$ backbone evaluations, whereas Heun's method requires $2S-1$ evaluations when the final correction step is omitted~\cite{2008_Butcher_Book,2022_Karras_NeurIPS_35}. 
Accordingly, backbone Gflops alone do not fully determine inference cost, and comparisons between published models should also take solver dependent evaluation counts into account.

\section{CCD2 Experiments and Results}
\label{sec:ccd2_results}

\begin{table}[t]
\centering
\small
\setlength{\tabcolsep}{5pt}
\resizebox{\columnwidth}{!}{%
\begin{tabular}{lccccc}
\toprule
& \multicolumn{3}{c}{Classifier AUC} & & \\
\cmidrule(lr){2-4}
Model & High-level & Low-level & ResNet & FPD $\times 10^3$ & Gen.\ time [ms/shower] \\
\midrule
Geant4 & $0.499 \pm 0.002$ & $0.500 \pm 0.002$ & $0.500 \pm 0.004$ & $10.7 \pm 0.8$ & -- \\
\midrule
CaloDiT-2 CD & $0.598 \pm 0.002$ & $0.695 \pm 0.004$ & -- & $68.83 \pm 3.37$ & -- \\
CaloDiT-2 EDM & $0.560 \pm 0.005$ & $0.594 \pm 0.002$ & -- & $20.06 \pm 0.74$ & -- \\
\midrule
CaloDREAM & $0.521 \pm 0.002$ & $0.531 \pm 0.003$ & $0.681 \pm 0.015$ & $25.0 \pm 1.0$ & $74.3 \pm 0.8$ \\
CaloDREAM++ & $0.512 \pm 0.001$ & $\mathbf{0.516 \pm 0.001}$ & $0.683 \pm 0.009$ & $16.0 \pm 0.5$ & $34 \pm 1$ \\
\midrule
CaloArt & $\mathbf{0.508 \pm 0.003}$ & $0.526 \pm 0.003$ & $\mathbf{0.632 \pm 0.068}$ & $\mathbf{14.11 \pm 0.68}$ & $9.71$ \\
\bottomrule
\end{tabular}%
}
\caption{Comparison on CCD2 between some published transformer backbone models and our CaloArt result. Geant4 is shown as a reference.}
\label{tab:ccd2_baselines_caloart}
\end{table}

Experiments in this work are carried out on CaloChallenge Dataset-2 (CCD2) and Dataset-3 (CCD3), 
two public regular-grid calorimeter benchmarks from the Fast Calorimeter Simulation Challenge that have become established testbeds for comparing generative calorimeter models across different formulations. 
In this section, we focus on CCD2, where the energy in the absorber layers is voxelized into a regular grid with $45$ longitudinal layers and a $16\times 9$ segmentation in the angular $\alpha$ and radial $r$ directions. 
The minimum readout energy, which is also used as the threshold for sparsity, is $x_{\mathrm{th}}=15.15\,\mathrm{keV}$. 
The shower is conditioned only on the incident energy $E_{\mathrm{inc}}$, 
which is sampled log-uniformly over $1\text{--}1000\,\mathrm{GeV}$.

Figure~\ref{fig:xpredcompare} shows that, in the CCD2 setting considered here, 
$\vv$-prediction gives stronger overall performance than $\x$-prediction under the patch size setting used here. Unless otherwise stated,
all CCD2 experiments in this section therefore use $\vv$-prediction.

\subsection{Main results on CCD2}
\label{subsec:ccd2_main_results}

Table~\ref{tab:ccd2_baselines_caloart} summarizes the main CCD2 result of CaloArt together with several published transformer backbone reference models. 
CaloDiT-2 EDM is included as a direct full-shower diffusion baseline, 
while CaloDREAM and CaloDREAM++ provide strong transformer based flow references. 

Overall, CaloArt is strongly competitive with these published models on CCD2. 
Its clearest advantage appears in FPD, where it achieves the lowest value among the learned models in this comparison, improving over both CaloDiT-2 EDM and the CaloDREAM-family results. 
At the same time, CaloArt gives the best high-level AUC and the strongest ResNet AUC in the table, while CaloDREAM++ remains slightly better on the low-level classifier. 
Generation time is also an important consideration for fast calorimeter simulation. 
The CaloDiT-2 paper reports batch size 1 timings for its EDM and consistency-distilled variants, 
but does not provide the corresponding batch size 100 timing used here. 
The distilled CaloDiT-2 CD model requires only a single network evaluation and is therefore expected to be the fastest. 
Among the non-distilled diffusion and flow models with reported timings, CaloArt is faster than both CaloDiT-2 EDM and CaloDREAM++. 
In the following subsections, we analyze this generation-time behavior in terms of backbone Gflops and solver dependent evaluation counts, 
which together determine the total inference compute. 

\subsection{Increasing forward pass Gflops improves shower sample quality}

A central conclusion of DiT~\cite{2022_Peebles_arXiv_2212_09748} is that increasing transformer forward pass Gflops improves sample quality. 
In DiT, this increase can come either from scaling the transformer backbone, making it deeper and wider, 
or from using finer patch tokenization, where decreasing the patch size increases the number of input tokens. 
Motivated by this perspective, we perform a smaller CCD2 sweep over backbone scale and patch size for direct calorimeter shower transformer backbones. 
Owing to limited computational resources, the experiments here are restricted to seven runs, 
with the maximum backbone compute capped at $4.431$ Gflops, matching the scale of the released CaloDiT-2 backbone.

\begin{figure}[t]
\centering
\includegraphics[width=.80\columnwidth]{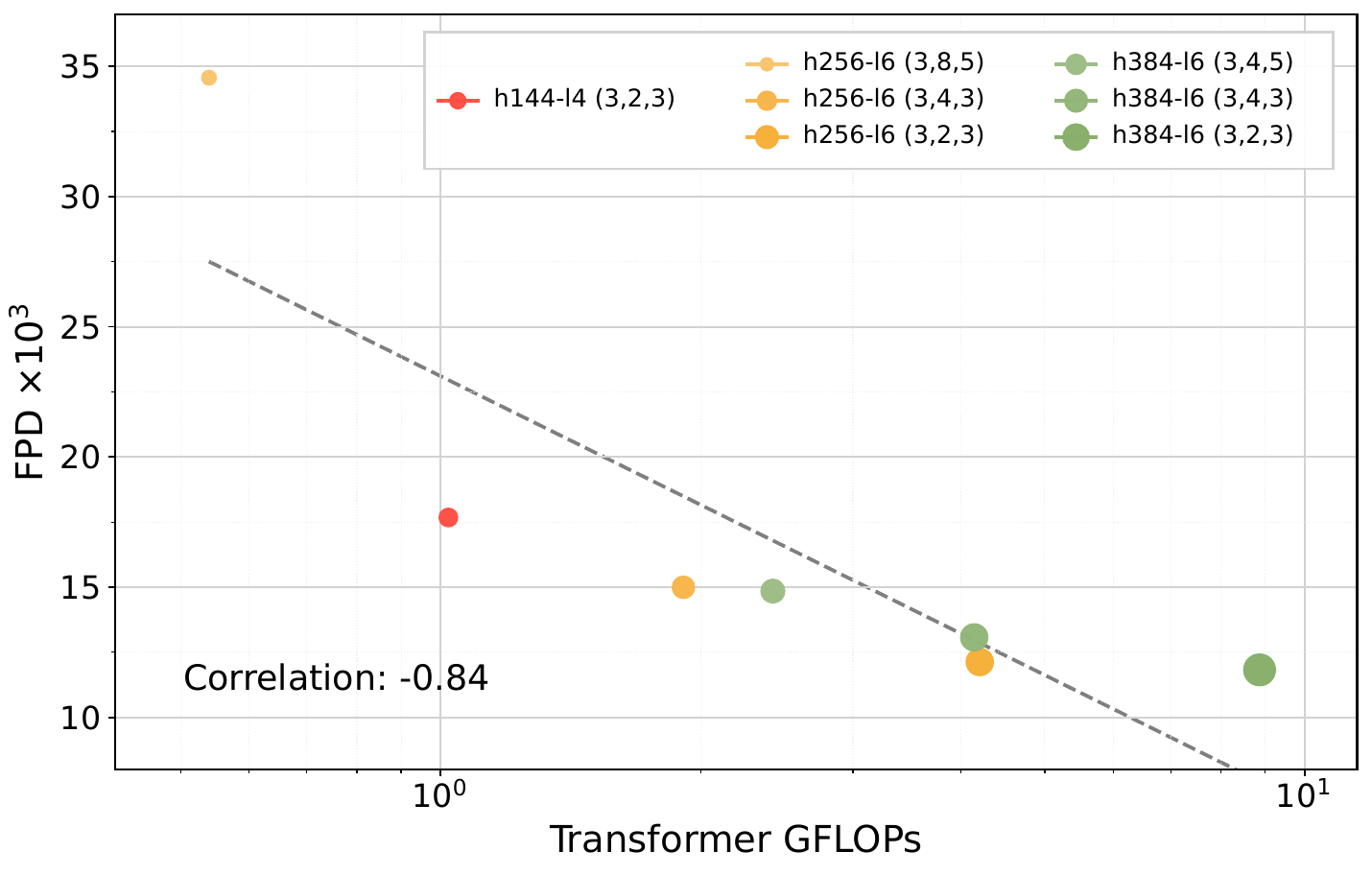}
\caption{
\textbf{FPD decreases as transformer backbone Gflops increase on CCD2.} Each point shows one of our seven CCD2 DiT models, using its backbone Gflops and the lowest FPD reached within 400K training steps. The trend follows the DiT compute--quality relation: higher backbone compute, from either scaling the backbone or using finer patches, gives better sample quality.
}
\label{fig:ccd2_fpd_Gflops_correlation}
\end{figure}

Figure~\ref{fig:ccd2_fpd_Gflops_correlation} shows the same qualitative trend on CCD2: as backbone Gflops increase, FPD decreases substantially, 
indicating better agreement with the Geant4 reference in high-level shower features. 
This improvement is obtained both by scaling the backbone and by decreasing patch size, consistent with the original DiT observation 
that model compute, rather than parameter count alone, is the more informative organizing axis for transformer diffusion model performance. 
As in DiT, different model configurations achieve similar sample quality at similar Gflops; for example \texttt{h256\_l6} with patch size $(3,2,3)$ and \texttt{h384\_l6} with patch size $(3,4,3)$ are very close. 
Appendix~\ref{app:ccd2_cls_Gflops} further shows that this trend holds for the three classifier-based metrics.

\subsection{Gflops, model scale, and sampling design choices}
\label{subsec:Gflops_model_scale_sampling_choice}

\begin{table}[t]
\centering
\small
\setlength{\tabcolsep}{4pt}
\resizebox{\textwidth}{!}{%
\begin{tabular}{llcccccccc}
\toprule
Family & Dataset & Hidden dim & Layers & Patch sizes & Tokens & Params (M) & Solver & Steps & Gflops \\
\midrule
CaloDiT-2 EDM & CCD2 & 384 & 6 &$(3,2,3)$ & 360 & 17.842 & Heun & 32 & 4.437 \\
\midrule
CaloDREAM++ & CCD2 & 480 & 6 & $(3,16,1)$ & 135 & 26.043 & RK4 & 20 & 2.258 \\
CaloDREAM++ & CCD3 & 480 & 6 & $(3,10,3)$ & 450 & 26.083 & RK4 & 20 & 7.522 \\
\midrule
CaloArt & CCD2 & 384 & 6 & $(3,4,3)$ & 180 & 16.599 & Heun & 32 & 2.071 \\
CaloArt & CCD3 & 480 & 6 & $(5,10,6)$ & 135 & 18.765 & Heun & 32 & 2.386 \\
\bottomrule
\end{tabular}%
}
\caption{Model scale, tokenization, and sampling configurations for representative transformer-based diffusion and flow calorimeter generators. Reported Gflops follow the DiT-compatible forward-pass convention discussed above.}
\label{tab:Gflops_sample_compare_models_ours}
\end{table}

The scaling study above shows that increasing backbone Gflops substantially improves sample quality in the calorimeter DiT setting, 
consistent with the compute-scaling behavior observed in DiT~\cite{2022_Peebles_arXiv_2212_09748}. 
Since the goal of this work is to evaluate refinements to the calorimeter DiT backbone, 
rather than to obtain improvements simply from higher forward pass Gflops or from using more ODE solver evaluations at inference, 
we choose the final CaloArt configurations under computational constraints.

Such a compute budget constraint is directly relevant when comparing published transformer calorimeter generator baselines. 
The released non-distilled CaloDiT-2 model uses a $32$-step Heun sampler, corresponding to $63$ backbone evaluations, 
with a per forward complexity of $4.437$ Gflops~\cite{2025_Raikwar_arXiv_2509_07700}. 
In contrast, the CCD2 CaloDREAM++ shape model uses a $20$-step RK4 solver, corresponding to $80$ backbone evaluations, 
with a lower per forward complexity of $2.258$ Gflops~\cite{Favaro:2026awn}. 
Combining the per forward Gflops with the solver dependent number of evaluations gives total inference computational costs of approximately $4.437 \times 63$ for CaloDiT-2 and $2.258 \times 80$ for CaloDREAM++. 
Thus, although these two baselines differ in single-forward backbone complexity, 
their total inference computational costs on CCD2 are already of the same order. 
This makes them useful reference points for choosing a compute-constrained CaloArt configuration.

On CCD2, the final CaloArt configuration is chosen to stay within the lower end of the published baseline range in terms of backbone size, forward pass Gflops, and solver evaluation count. 
In terms of backbone size, CaloArt uses the same hidden dimension and depth as the released CaloDiT-2 model, 
with hidden dimension $384$ and $6$ transformer layers, which is the smallest backbone scale considered here~\cite{2025_Raikwar_arXiv_2509_07700}. 
The patch size is then set to $(3,4,3)$ instead of the CaloDiT-2 choice $(3,2,3)$, 
reducing the token count from $360$ to $180$. 
This lowers the single forward complexity from the CaloDiT-2 scale of $4.437$ Gflops to $2.071$ Gflops,
slightly below the $2.258$ Gflops of the CCD2 CaloDREAM++ shape model~\cite{Favaro:2026awn}. 
CaloArt uses the same $32$-step Heun solver as the non-distilled CaloDiT-2 setting, corresponding to $2S-1=63$ backbone evaluations, 
fewer than the $80$ backbone evaluations used by CaloDREAM++, which adopts a $20$-step RK4 solver~\cite{2025_Raikwar_arXiv_2509_07700,Favaro:2026awn}. 
Taken together, the lighter forward pass and the smaller solver-evaluation count give CaloArt a lower theoretical inference cost than CaloDREAM++ on CCD2, 
which accounts for part of the generation time advantage in Table~\ref{tab:ccd2_baselines_caloart}, 
the remaining speedup comes from implementation-level differences.

\subsection{Effect of introducing RoPE}
\label{subsec:RoPE_ablation}

\begin{table}[t]
\centering
\small
\setlength{\tabcolsep}{4pt}
\begin{tabular}{lcccc}
\toprule
& \multicolumn{3}{c}{Classifier AUC} & \\
\cmidrule(lr){2-4}
Positional encoding & High-level & Low-level & ResNet & FPD $\times 10^3$ \\
\midrule
\texttt{ape-only}  & $0.509 \pm 0.003$ & $0.534 \pm 0.003$ & $0.731 \pm 0.007$ & $14.19 \pm 0.82$ \\
\texttt{rope-only} & $\mathbf{0.506 \pm 0.003}$ & $\mathbf{0.523 \pm 0.003}$ & $0.716 \pm 0.012$ & $13.99 \pm 0.81$ \\
\texttt{ape+rope}  & $0.508 \pm 0.003$ & $0.528 \pm 0.003$ & $\mathbf{0.710 \pm 0.039}$ & $\mathbf{13.94 \pm 0.95}$ \\
\bottomrule
\end{tabular}%
\caption{CCD2 positional-encoding ablation under matched CaloArt configurations.}
\label{tab:ccd2_posenc_ablation}
\end{table}

Following the model scale comparison in Table~\ref{tab:Gflops_sample_compare_models_ours}, we perform most experiments on CCD2, 
where CaloArt uses a lighter backbone and lower Gflops. The lower-resolution voxel grid also makes repeated training, sample generation, and evaluation faster, while reducing storage and I/O pressure. 
The positional-encoding ablation is therefore carried out on CCD2. 
Table~\ref{tab:ccd2_posenc_ablation} compares the three position encoding variants \texttt{ape-only}, \texttt{rope-only}, and \texttt{ape+rope}.  
Introducing RoPE~\cite{2021_Su_RoFormer} leads to a clear improvement over using absolute position encoding alone, 
which is common in ViT/DiT-style backbones~\cite{2021_Dosovitskiy_ICLR,2022_Peebles_arXiv_2212_09748}. 
Both \texttt{rope-only} and \texttt{ape+rope} improve FPD together with the high-level classifier results, indicating better agreement in global shower statistics and high-level observables. 
The low-level and ResNet-based classifier results are also improved, indicating higher fidelity at the level of individual generated showers.

At the same time, \texttt{rope-only} and \texttt{ape+rope} remain close overall on this task, 
suggesting that once RoPE is introduced, adding APE together does not provide a clear additional gain in the present setting. 
Nevertheless, we retain \texttt{ape+rope} as the default positional design in CaloArt. 
In addition to giving a slightly stronger overall result in this comparison, 
this choice is also consistent with DiT-style upgrades that introduce RoPE while retaining an absolute positional component~\cite{2025_LightningDiT}, 
and remains more convenient for architectural modifications such as in-context class conditioning, 
where repeated conditioning tokens are assigned distinct absolute positional embeddings~\cite{2024_Li_NeurIPS_MAR,2025_Li_arXiv_2511_13720}.

\subsection{\texorpdfstring{$\x$-prediction versus $\vv$-prediction on CCD2}{x-prediction versus v-prediction on CCD2}}

\label{subsec:xpred_vpred_early}

Besides backbone refinement, a central theme of this work is whether an $\x$-prediction formulation is useful for diffusion calorimeter simulation. 
This depends on both detector resolution and patch granularity. 
On CCD2, the shower tensor contains $45\times16\times9=6480$ voxels, placing it in a comparatively low-resolution regime.
As a reference, this dimensionality is close to common image latent-diffusion dimensions, such as $4\times32\times32$ and $4\times64\times64$ under an $f=8$ autoencoder~\cite{2022_Rombach_CVPR_LDM,2022_Peebles_arXiv_2212_09748}. 
In this latent regime, transformer diffusion models commonly use small patch sizes, such as $2\times2$ or $4\times4$, 
while both noise prediction diffusion models and velocity prediction flow models are widely used modeling choices~\cite{2021_Song_ICLR_DDIM,2024_Ma_ECCV}. 
Although CaloArt operates directly on physical voxels rather than on a learned image latent space, this comparison provides a useful reference.

\begin{figure}[t]
\centering
\includegraphics[width=0.9\textwidth]{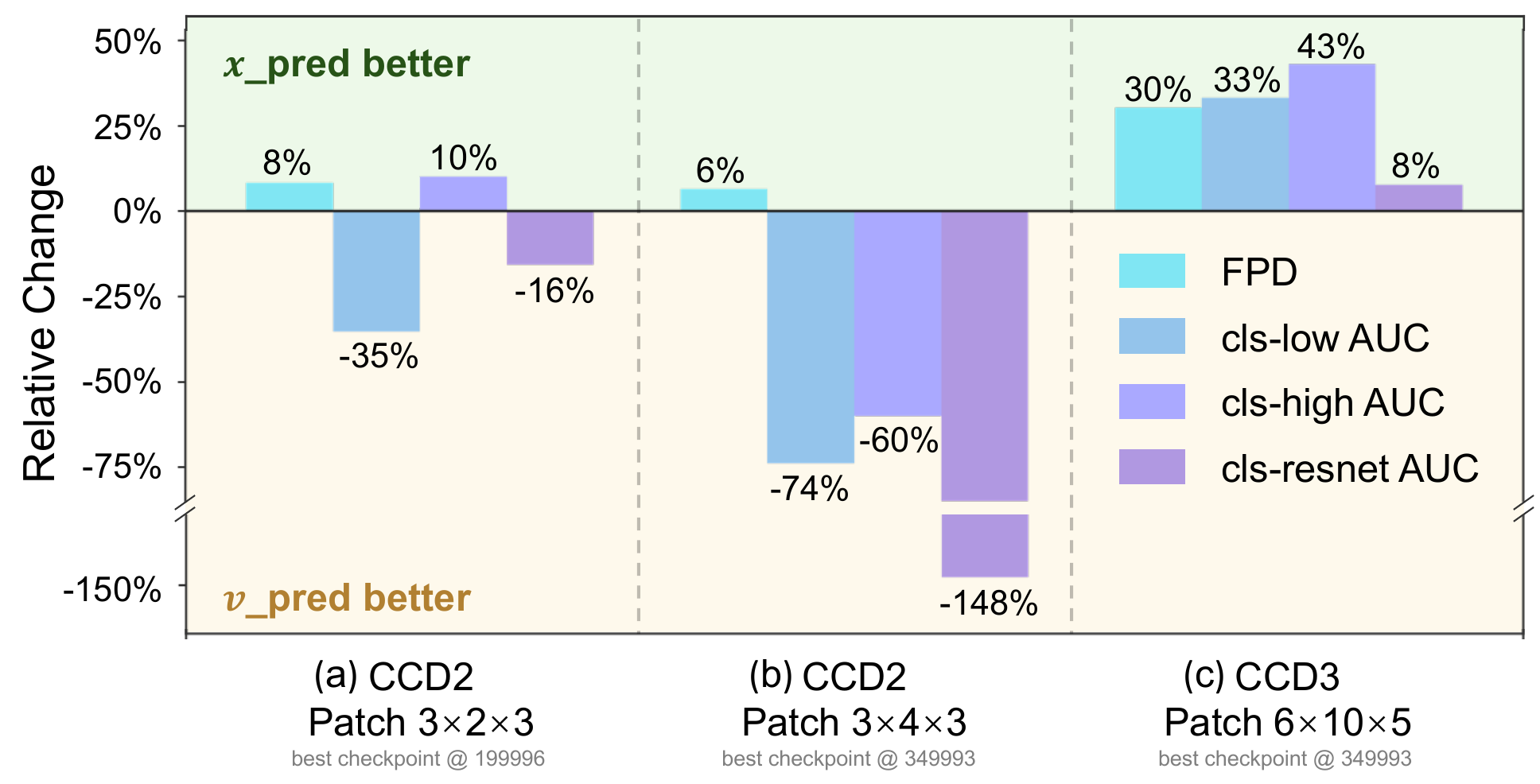}
\caption{\textbf{Relative change from $\vv$-prediction to $\x$-prediction in three representative configurations.} 
Each panel reports FPD and three classifier AUC metrics. 
For FPD, the value is $(x_{\mathrm{pred}}-v_{\mathrm{pred}})/v_{\mathrm{pred}}$; 
for AUCs, it is the relative change in distance from $0.5$, measuring classifier distinguishability. 
Negative values favor $\x$-prediction, while positive values favor $\vv$-prediction. 
Panels (a) and (b) show two CCD2 patch-size settings with mixed metric-level effects, 
whereas panel (c) shows a representative CCD3 setting where $\x$-prediction improves all metrics.}
\label{fig:xpredcompare}
\end{figure}

Figure~\ref{fig:xpredcompare} summarizes the relative change obtained by replacing $\vv$-prediction with $\x$-prediction under representative model configurations used in this paper. 
Negative values indicate an improvement when using $\x$-prediction.
On CCD2, the two tested patch size settings do not show a uniform advantage for $\x$-prediction. 
Instead, $\x$-prediction improves some classifier-based metrics in one setting, but also slightly worsens FPD. 
More importantly, for the final CCD2 patch size $(3,4,3)$, the classifier advantage of $\vv$-prediction is larger than the FPD gain obtained from switching to $\x$-prediction. 
We therefore use $\vv$-prediction as the default CCD2 formulation.

This choice should be considered together with the compute setting discussed above. 
Under the CCD2 Gflops budget, the model can still use a relatively small patch size without excessive token number pressure. 
The final CCD2 recipe therefore combines sufficiently fine patch granularity with $\vv$-prediction.

\section{CCD3 Experiments and Results}
\label{sec:ccd3_results}

We now move to CCD3, where the much finer detector granularity defines a regime qualitatively different from CCD2.
As in CCD2, the energy in the absorber layers is voxelized into $45$ longitudinal layers, 
but the transverse segmentation is now $50\times18$ in the angular $\alpha$ and radial $r$ directions, 
corresponding to a total shower shape of $45\times50\times18$. Relative to CCD2, 
this substantially higher resolution imposes much stronger token count pressure on direct full-shower transformer models. 
Figure~\ref{fig:xpredcompare} highlights the difference between prediction targets in this regime:
in the rightmost CCD3 panel, under the patch size setting considered here, $\x$-prediction improves over $\vv$-prediction on all displayed metrics rather than only in the overall trade-off.

\begin{table}[t]
\centering
\small
\setlength{\tabcolsep}{5pt}
\resizebox{\columnwidth}{!}{%
\begin{tabular}{llccccc}
\toprule
& & \multicolumn{3}{c}{Classifier AUC} & & \\
\cmidrule(lr){3-5}
Model & Dataset & High-level & Low-level & ResNet & FPD $\times 10^3$ & Gen.\ time [ms/shower] \\
\midrule
CaloDREAM & CCD3 & $\mathbf{0.524 \pm 0.004}$ & $0.630 \pm 0.005$ & $0.802 \pm 0.014$ & $\mathbf{20.7 \pm 1.1}$ & $179.6 \pm 0.5$ \\
CaloDREAM++ & CCD3 & $0.538 \pm 0.002$ & $\mathbf{0.524 \pm 0.001}$ & $\mathbf{0.799 \pm 0.009}$ & $26.3 \pm 0.4$ & $96 \pm 1$ \\
CaloArt & CCD3 & $0.566 \pm 0.004$ & $0.579 \pm 0.004$ & $0.8350 \pm 0.009$ & $42.2 \pm 1.0$ & $\mathbf{11.14}$ \\
conv.\ L2LFlows & CCD3 & $0.733 \pm 0.006$ & $0.588 \pm 0.004$ & $0.919 \pm 0.003$ & $171.6 \pm 1.8$ & $16.0 \pm 0.2$ \\

\bottomrule
\end{tabular}%
}
\caption{Comparison on CCD3 between published CaloDREAM-family transformer flow models, a convolutional L2LFlows reference, and our CaloArt result.}
\label{tab:ccd3_calodream_caloart}
\end{table}

\subsection{Main results on CCD3}
\label{subsec:ccd3_main_results}

Table~\ref{tab:ccd3_calodream_caloart} summarizes the main CCD3 result of CaloArt together with three published reference points for CCD3~\cite{2025_Krause_RPP_88}. 
CaloDREAM and CaloDREAM++ are included as strong transformer-based flow models on CCD3~\cite{Favaro:2024rle,Favaro:2026awn}, 
while convolutional L2LFlows is included as a non-transformer reference with a comparable reported generation time. 
The CaloDREAM family still gives the best sample quality, with lower FPD values and stronger classifier-based metrics. 
Nevertheless, CaloArt remains competitive on both FPD and classifier AUCs, while achieving this level of quality with a much shorter reported GPU generation time. 
The comparison with convolutional L2LFlows makes the quality--speed balance more explicit. 
At a comparable reported generation time, CaloArt achieves substantially better FPD and stronger classifier AUCs, while also being faster in the reported timing. 

\begin{figure}[t]
\centering
\includegraphics[width=0.85\columnwidth]{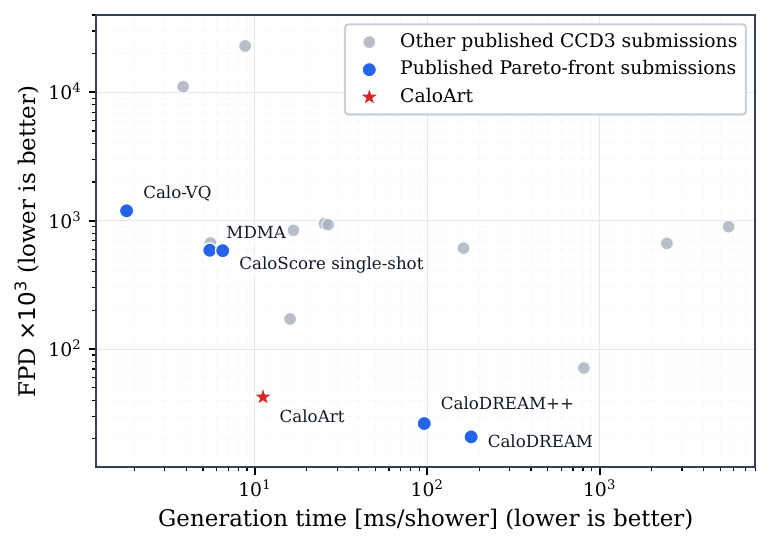}
\caption{\textbf{FPD--generation-time Pareto view on CCD3.} CCD3 methods with reported GPU generation time are compared using time per shower and FPD $\times 10^3$, with lower-left preferred. Blue points mark published Pareto-front submissions, and the red star marks CaloArt, which remains Pareto-efficient in this comparison.}
\label{fig:ccd3_fpd_time_pareto}
\end{figure}

To further substantiate this quality--speed trade-off, 
Figure~\ref{fig:ccd3_fpd_time_pareto} plots the broader published CCD3 landscape in the FPD--generation-time plane. 
The data are collected in Appendix~\ref{app:ccd3_public_submissions} from the published CCD3 summary tables~\cite{2025_Krause_RPP_88}. 
Since lower values are preferred for both objectives, the lower-left envelope defines the empirical Pareto frontier of the timed methods considered here~\cite{1999_Miettinen_Book}. CaloArt lies on the empirical Pareto frontier: lower-FPD methods are substantially slower, while faster methods have much larger FPD.

\subsection{\texorpdfstring{CCD3 as a well-suited regime for $\x$-prediction}{CCD3 as a well-suited regime for x-prediction}}

CCD3 places the prediction target question in a substantially different regime. 
Its shower has a total shape of $45\times50\times18=40500$ voxels, 
placing direct full-shower generation much closer to raw pixel image modeling than to the latent-space diffusion regime discussed above~\cite{2022_Rombach_CVPR_LDM}. 
For reference, this dimensionality is of the same order as common raw-pixel inputs such as a $128\times128$ RGB image~\cite{2021_Dhariwal_NeurIPS_ADM}, 
while the shower itself remains highly sparse.

The latent tokenizer route~\cite{2021_Esser_CVPR_Taming,2022_Rombach_CVPR_LDM} is difficult to follow in this setting. 
For calorimeter showers, there is no broadly established counterpart of the pre-trained VGG-style classifiers~\cite{2014_Simonyan_arXiv_VGG} or DINO-style self-supervised feature extractors~\cite{2023_Oquab_arXiv_DINOv2}. 
These visual representations are used to define perceptual losses~\cite{2016_Johnson_ECCV_PercepLoss,2018_Zhang_CVPR_LPIPS} or representation-alignment objectives~\cite{2025_Yu_ICLR_REPA} when training image autoencoders and related latent generative models, 
but no standard representation exists for shower data. 
As a result, a VAE-like tokenizer may compress the voxel grid while failing to preserve the physics observables needed for accurate calorimeter simulation. 
This concern is also reflected in CaloDiT-2, where VAE-like components are deliberately avoided because they lead to overly smoothed or blurry showers~\cite{2025_Raikwar_arXiv_2509_07700}.

At the same time, direct raw voxel generation on CCD3 is strongly constrained by both training and inference cost. 
This issue has also been emphasized in the CaloDREAM++ paper~\cite{Favaro:2026awn}: on Dataset-3, fully training the earlier configuration exceeded the available computational resources, 
so the authors enlarged the patch size to reduce the number of embedded patches and improve convergence under a feasible training budget. 
As shown in Table~\ref{tab:Gflops_sample_compare_models_ours}, even the CCD3 CaloDREAM++ shape model~\cite{Favaro:2026awn} uses a relatively large patch size, $(3,10,3)$, which reduces the sequence length to $450$ tokens but still gives $7.522$ Gflops per forward pass. 
CaloArt adopts a more aggressive patch size, $(5,10,6)$, reducing the sequence length to $135$ tokens and the forward complexity to $2.386$ Gflops. 
Thus, a large patch size is a necessary design choice on CCD3 under practical fast simulation constraints.

Taken together, these properties make CCD3 a calorimeter counterpart to the setting considered in JiT~\cite{2025_Li_arXiv_2511_13720}: 
high-dimensional raw data, a tokenizer that is difficult to design, and compute constraints that force the model to use relatively large patches. 
From this perspective, CCD3 is a regime where directly predicting the clean sample, $\x$-prediction~\cite{2025_Li_arXiv_2511_13720}, should be a good fit.

\subsection{\texorpdfstring{Clear gains of $\x$-prediction on CCD3}{Clear gains of x-prediction on CCD3}}

Figure~\ref{fig:xpredcompare} summarizes the effect of switching from $\vv$-prediction to $\x$-prediction. 
While $\x$-prediction does not provide an advantage on CCD2, the CCD3 result is qualitatively different: 
in panel~(c), under the CCD3 patch size setting used for the final model, 
switching to $\x$-prediction leads to a consistent improvement across all reported metrics, 
lowering FPD and moving the classifier AUCs for the high-level, low-level, and ResNet evaluators closer to the ideal value of $0.5$. 
This suggests that, in a higher-resolution calorimeter simulation setting where substantially coarser tokenization is needed to control the token count and backbone Gflops, $\x$-prediction can improve metrics without an evident trade-off.

\begin{figure}[t]
\centering
\includegraphics[width=0.82\linewidth]{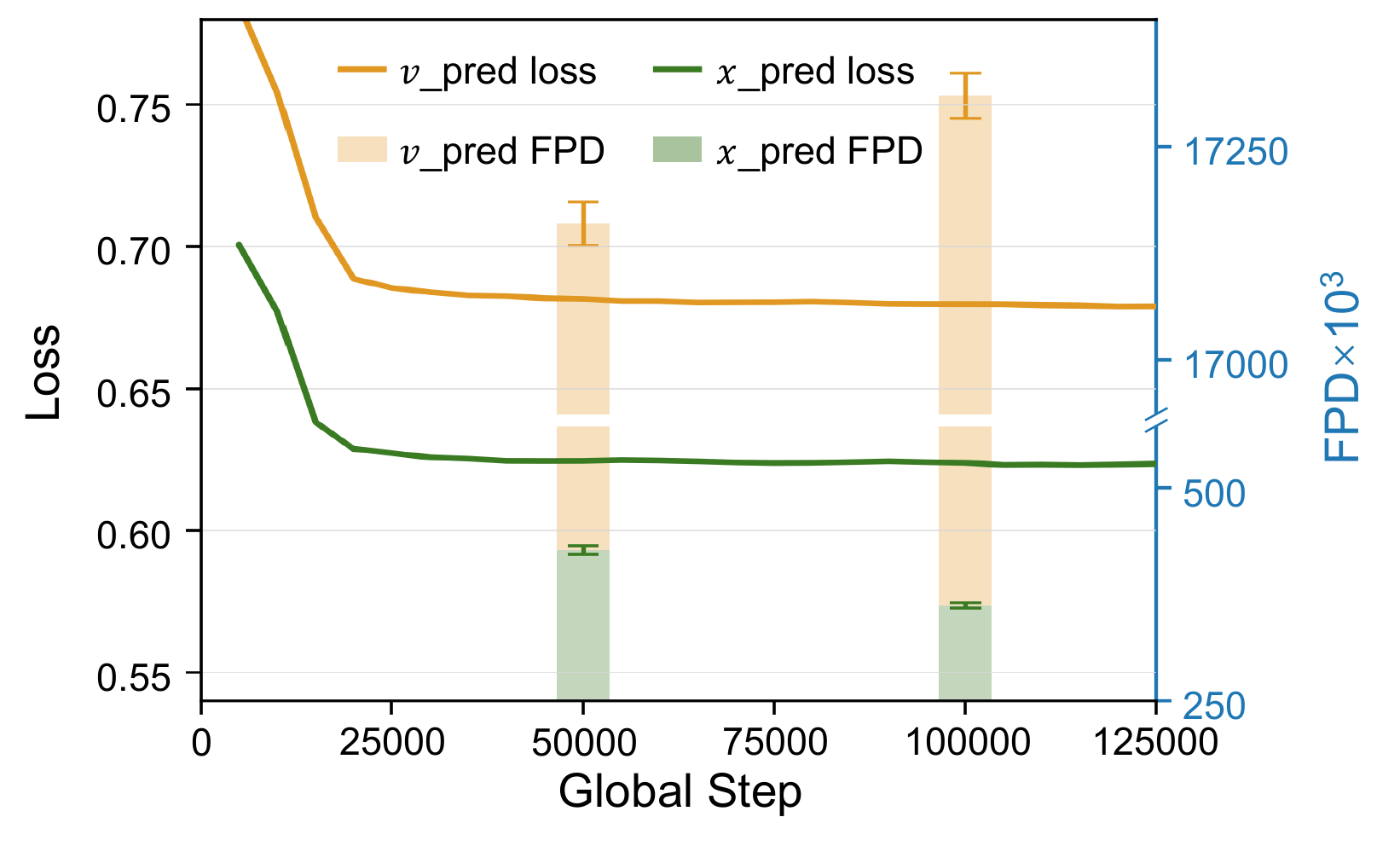}
\caption{\textbf{Validation-loss comparison between $\x$-prediction and $\vv$-prediction on CCD3.} Both curves are shown against \texttt{global\_step} for the same setting. The callouts report the FPD values at the $50$K- and $100$K-step checkpoints for both formulations.}
\label{fig:ccd3_loss_xpred_vpred}
\end{figure}

The $\vv$-prediction behavior in Figure~\ref{fig:xpredcompare} panel~(c) is less extreme than the collapse reported in the JiT paper for large-patch image generation, 
where $\x$-prediction can be trained successfully while $\vv$-prediction fails to produce meaningful samples~\cite{2025_Li_arXiv_2511_13720}. 
This difference is plausible because, although the $(5,10,6)$ patch size is large and the hidden dimension $480$ is smaller than the image DiT settings considered in JiT, 
the effective dimensionality of the calorimeter shower remains lower than that of high-resolution raw images~\cite{2006_Chapelle_Book}. 
A catastrophic failure of $\vv$-prediction appears once the patch size is made even more aggressive. 
Figure~\ref{fig:ccd3_loss_xpred_vpred} compares the training loss curves under the same $\vv$-loss, 
using $\vv$-prediction versus $\x$-prediction for a hidden-$384$ CCD3 setting with patch size $(5,10,9)$. 
Since the loss is computed in the same space and only the prediction parameterization differs, comparing the loss values is legitimate. 
The $\x$-prediction curve remains consistently below the corresponding $\vv$-prediction curve, 
so the failure of $\vv$-prediction is already visible in the training loss curves at the optimization level. 
For $\vv$-prediction, the FPD stays in an anomalously high regime, remaining around \(1.72\times 10^{4}\) at both reported checkpoints and even increasing slightly at the later checkpoint. 
This indicates that the run remains stuck far from a usable low FPD model, with generated samples that fail to match the reference shower distribution. 
In contrast, $\x$-prediction shows a normal improvement trajectory in FPD for this setting, 
decreasing within the few-hundred range from about \(427\) to \(362\) over the same checkpoints. 
Thus, under the more aggressive patch size, only $\x$-prediction remains trainable in a practical sense.

\subsection{Shared modulation}
\label{subsec:shared_modulation_ablation}

\begin{table}[t]
\centering
\footnotesize
\setlength{\tabcolsep}{2pt}
\renewcommand{\arraystretch}{1.0}
\resizebox{0.92\columnwidth}{!}{%
\begin{tabular}{@{}ccccccc@{}}
\toprule
& & & \multicolumn{3}{c}{Classifier AUC} & \\
\cmidrule(lr){4-6}
Shared modulation & Params (M) & Gflops & Low-level & High-level & ResNet & FPD $\times 10^3$ \\
\midrule
Yes & $\mathbf{18.765}$ & $\mathbf{2.386}$ & $\mathbf{0.579 \pm 0.004}$ & $\mathbf{0.566 \pm 0.004}$ & $\mathbf{0.835 \pm 0.009}$ & $42.23 \pm 1.01$ \\
No  & 26.135 & 2.393 & $0.637 \pm 0.005$ & $0.572 \pm 0.002$ & $0.841 \pm 0.005$ & $\mathbf{41.03 \pm 1.09}$ \\
\bottomrule
\end{tabular}
}
\caption{CCD3 ablation of shared modulation under matched CaloArt configurations.}
\label{tab:ccd3_sharemod_ablation}
\end{table}

By making larger patch sizes viable on CCD3, $\x$-prediction helps keep the token count and forward Gflops manageable, 
leaving more room to scale the backbone width and depth. 
Such backbone scaling increases the parameter count. 
We therefore examine shared modulation as a lightweight alternative to the standard block-wise DiT conditioning branch~\cite{2022_Peebles_arXiv_2212_09748}, 
aiming to shift parameter budget from repeated adaptive-normalization projections toward the backbone itself~\cite{2024_Gupta_WALT}.

This ablation follows a broader line of parameter efficient adaptive-normalization designs in diffusion transformers. 
PixArt-$\alpha$~\cite{2024_Chen_PixArtAlpha} introduced \textit{adaLN-single} as a lighter alternative to the standard adaptive-normalization pathway, 
while DiT-Air~\cite{2025_Chen_DiTAir} further explored shared-AdaLN designs for improving parameter efficiency in diffusion transformers. 
In PixArt-$\alpha$, \textit{adaLN-single} was introduced for shared timestep conditioning, while text conditions are handled by cross-attention. 
Its multi-scale implementation~\cite{2024_Chen_PixArtAlpha} further folds global size-related conditions, such as image size and aspect ratio, into the same modulation pathway, 
in line with SDXL-style conditioning~\cite{2023_Podell_SDXL}. 
This motivates our use of a shared modulation branch for the combined physical and time conditioning signal.

To make the shared modulation design explicit, 
we follow the \textit{adaLN-single} notation of PixArt-$\alpha$~\cite{2024_Chen_PixArtAlpha} 
and denote by \(S^{(i)} = [\beta_1^{(i)}, \beta_2^{(i)}, \gamma_1^{(i)}, \gamma_2^{(i)}, \alpha_1^{(i)}, \alpha_2^{(i)}]\) the full tuple of adaptive shift, scale, and gating parameters in the adaLN module of the \(i\)-th transformer block. 
Here the subscripts \(1\) and \(2\) refer to the modulation parameters used in the attention or FFN sublayers. 
In the standard DiT-style design~\cite{2022_Peebles_arXiv_2212_09748}, 
each block obtains its modulation parameters from a block-specific projection, \(S^{(i)} = f^{(i)}(c+t)\), where \(c\) denotes the physical-conditioning embedding and \(t\) denotes the timestep embedding.

In the PixArt-style shared design~\cite{2024_Chen_PixArtAlpha} used by CaloArt, 
one global modulation tuple \(\bar{S}=f(c+t)\) is first computed by a projection MLP and is then reused across all blocks. 
The block-specific modulation \(S^{(i)}\) is then obtained as \(S^{(i)}=g(\bar{S},E^{(i)})\), 
where \(g\) is a summation function and \(E^{(i)}\) is a layer-specific trainable embedding with the same shape as \(\bar{S}\). 
Equivalently, \(S^{(i)}=\bar{S}+E^{(i)}\), so the shared conditioning modulation is adaptively adjusted for different transformer blocks. 
This is the quantity shown schematically in Fig.~\ref{fig:caloart_backbone} as \(g(E_i,\bar{\beta},\bar{\gamma})\) or \(g(E_i,\bar{\alpha})\).

We perform an ablation on CCD3 to isolate the effect of shared modulation, 
with the two runs in Table~\ref{tab:ccd3_sharemod_ablation} differing only in whether the modulation branch is shared across transformer blocks. 
The shared modulation variant reduces the parameter count from $26.13$M to $18.77$M, corresponding to a reduction of about $28\%$ relative to the non-shared baseline. 
The forward Gflops change only marginally, from $2.393$ to $2.386$, 
because the removed block-wise modulation projection linear layers contain many parameters but contribute little to the overall arithmetic cost. 
In terms of sample quality, the two variants remain close. The non-shared baseline gives a slightly lower FPD, $41.03 \pm 1.09$, compared with $42.23 \pm 1.01$ for the shared-modulation model, 
while shared modulation improves all classifier-based metrics. 
These results suggest that shared modulation removes a substantial number of conditioning-branch parameters without introducing a clear degradation in overall generation quality. 

\section{Conclusion}
\label{sec:Conclusion}

We presented CaloArt as a refined DiT-style backbone for direct raw voxel  
calorimeter shower generation. The model keeps the end-to-end full shower  
formulation, without introducing a separate autoencoder tokenizer or an  
explicit layer-wise energy factorization. CaloArt modernizes this direct  
backbone with 3D positional encoding, RMSNorm, qk-norm and shared conditioning modulation.  In this sense, CaloArt is intended to serve  
as a stronger default DiT backbone for voxel-based calorimeter generation.

The Gflops analysis further provides a DiT-aligned scale measure for comparing
transformer based calorimeter generators against the DiT-style models commonly used in image generation. Under this accounting, the
scaling study shows a clear trend: increasing forward pass Gflops, either
through finer patch tokenization or through backbone scaling, improves the
shower generation metrics. This observation motivates a compact CCD2
configuration of CaloArt, using one of the lowest parameter and lowest Gflops settings among the compared transformer backbone calorimeter models.
Even under this compact configuration, CaloArt performs strongly on CCD2, 
demonstrating the effectiveness of the refined backbone by achieving the best FPD in the comparison, 
together with the strongest high-level and ResNet classifier AUCs.

We also explored the prediction target of the diffusion model and successfully
adopted $\x$-prediction on CCD3, yielding a CaloArt
configuration that lies on the quality--generation-time Pareto frontier.
$\x$-prediction is especially relevant for high-granularity calorimeters,
where the high dimensional voxel grid puts direct raw voxel generation under strong
pressure from both training and inference costs. Under these constraints,
large-patch tokenization becomes a necessary design choice, and thus the
prediction target becomes critical for determining whether such large-patch
models can be trained reliably. Under the final CCD3 patch-size setting,
switching to $\x$-prediction brings clear improvements across all reported
metrics. The more aggressive patch-size study further shows that, under tighter computational constraints, 
$\x$-prediction remains practically trainable, whereas $\vv$-prediction fails to generate usable samples.

\paragraph{Code and data availability.}
The code is publicly available at \url{https://github.com/miemiemi/CaloArt}.
The generated samples and the complete set of high-level features used during evaluation are available at
\href{https://doi.org/10.5281/zenodo.19881295}{\texttt{https://doi.org/10.5281/zenodo.19881295}}.

\clearpage

\acknowledgments

This work used public GPU computing resources provided by the IHEP Computing
Center (IHEP CC), Institute of High Energy Physics, Chinese Academy of Sciences.
We thank IHEP CC for maintaining the computing platform.

\paragraph{Funding information}
This work was supported by the National Natural Science Foundation
of China (under Grant No. 11775249).

\phantomsection
\addcontentsline{toc}{section}{Appendix}
\addtocontents{toc}{\protect\setcounter{tocdepth}{0}}
\appendix

\begin{table}[h]
\centering
\resizebox{0.90\textwidth}{!}{%
\tablestyle{4pt}{1.02}
\begin{tabular}{l | x{96}x{96}}
 & \textbf{CaloArt-CCD2} & \textbf{CaloArt-CCD3} \\
\shline
\rowcolor[gray]{0.9}\multicolumn{3}{l}{\textbf{architecture}} \\
\midline
detector grid \((z,\alpha,r)\) & $(45,16,9)$ & $(45,50,18)$ \\
patch size \((z,\alpha,r)\) & $(3,4,3)$ & $(5,10,6)$ \\
number of patch tokens & $180$ & $135$ \\
hidden dimension & $384$ & $480$ \\
depth & $6$ & $6$ \\
attention heads & $6$ & $8$ \\
MLP ratio & $4$ & $4$ \\
positional encoding & \multicolumn{2}{c}{\texttt{ape+rope}} \\
RoPE frequencies & \multicolumn{2}{c}{$(1.0,10000.0)$} \\
\midline
\rowcolor[gray]{0.9}\multicolumn{3}{l}{\textbf{training}} \\
\midline
prediction target & \(\vv\)-prediction & \(\x\)-prediction \\
loss target & \multicolumn{2}{c}{\(\vv\)-loss} \\
time sampler & $\mathrm{logit}(t)\sim\mathcal{N}(0.0,1.0^2)$ & $\mathrm{logit}(t)\sim\mathcal{N}(-0.5,1.0^2)$ \\
noise scale & \multicolumn{2}{c}{$1.0$} \\
optimizer & \multicolumn{2}{c}{AdamW, $\beta_1,\beta_2=0.9,0.95$, $\epsilon=10^{-8}$} \\
global batch size & \multicolumn{2}{c}{$256$} \\
training steps & \multicolumn{2}{c}{$400000$} \\
learning rate & $5.0\times10^{-4}$ & $8.0\times10^{-4}$ \\
learning-rate schedule & cosine & warmup--stable--decay \\
warmup steps & \multicolumn{2}{c}{$10000$} \\
stable / decay steps & -- & $180000 / 210000$ \\
minimum learning-rate ratio & -- & $0.1$ \\
weight decay & \multicolumn{2}{c}{$1.0\times10^{-5}$} \\
EMA schedule & \multicolumn{2}{c}{inverse decay, power $0.7$, max value $0.9999$} \\
\midline
\rowcolor[gray]{0.9}\multicolumn{3}{l}{\textbf{sampling}} \\
\midline
ODE solver & \multicolumn{2}{c}{Heun} \\
ODE steps & \multicolumn{2}{c}{$32$} \\
time steps & \multicolumn{2}{c}{linear in $[0.0,1.0]$} \\
reject-redraw sampling & no & yes, max ratio $2.7$ \\
\shline
\end{tabular}%
}
\vspace{-.5em}
\caption{\textbf{Configurations of the final CaloArt models.} The table reports the training configurations of the two final CCD2 and CCD3 models used for the main results.}
\label{tab:experiment_configurations}
\vspace{-.5em}
\end{table}

\section{Experimental Configurations}
\label{app:experimental_configurations}

Table~\ref{tab:experiment_configurations} summarizes the training configurations of the two final CaloArt models used for the CCD2 and CCD3 main results. 
The detector grid and patch size entries are reported in the paper convention \((z,\alpha,r)\), 
while the implementation configuration stores the corresponding tensor axes in the order used by the data loader.

\section{Prediction-Space and Loss-Space Combinations}
\label{app:xev}

This appendix summarizes the full set of prediction-space / loss-space combinations under the linear interpolation path used in Sec.~\ref{sec:method}. 
In the main text, we only discuss the two cases directly relevant to this work, 
$\vv$-prediction with $\vv$-loss and $\x$-prediction with $\vv$-loss. 
For completeness, Table~\ref{tab:xev_appendix} lists all nine combinations obtained by choosing the network prediction space and the loss space in \(x\), \(\epsilon\), or \(v\), 
together with the corresponding transformations.

At inference time, generation can always be carried out in the \(v\)-space: 
regardless of the chosen prediction target, 
the network output is first transformed to the corresponding \(v_\theta\), 
and the ODE in Eq.~\eqref{eq:sampling-ode} is then solved for sampling. 
As such, all nine combinations define valid generators.

\begin{table}[t]
\vspace{-1em}
\centering
{
\tablestyle{2.5pt}{1.2}
\small
\begin{tabular}{l l | x{90} | x{90} | x{90} }
& & \textbf{(a) \(x\)-pred } & \textbf{(b) \(\epsilon\)-pred} & \textbf{(c) \(v\)-pred} \\
& & \makebox[5em]{\(x_\theta := \mathrm{net}_\theta(z_t,t,c)\)} & \makebox[5em]{\(\epsilon_\theta := \mathrm{net}_\theta(z_t,t,c)\)} & \makebox[5em]{\(v_\theta := \mathrm{net}_\theta(z_t,t,c)\)} \\
\shline
\textbf{~(1)~\(x\)-loss:} & \(\mathbb{E}\|x_\theta - x\|^2\) &
\(x_\theta\) &
\(x_\theta=(z_t-(1-t)\epsilon_\theta) / t\) &
\(x_\theta=(1-t)v_\theta+z_t\) \\
\midline
\textbf{~(2)~\(\epsilon\)-loss:} & \(\mathbb{E}\|\epsilon_\theta - \epsilon\|^2\) &
\(\epsilon_\theta=(z_t-tx_\theta) / (1 - t)\) &
\(\epsilon_\theta\) &
\(\epsilon_\theta=z_t-tv_\theta\) \\
\midline
\textbf{~(3)~\(v\)-loss:} & \(\mathbb{E}\|v_\theta - v\|^2\) &
\(v_\theta=(x_\theta-z_t) / (1 - t)\) &
\(v_\theta=(z_t-\epsilon_\theta) / t\) &
\(v_\theta\) \\
\end{tabular}
}
\vspace{\baselineskip}
\caption{
\textbf{All possible combinations} of defining the loss and network prediction in \(x\), \(v\), or \(\epsilon\) spaces. 
The direct network outputs are shown explicitly in the column headers. 
For any off-diagonal entry where the network output space differs from the loss space, 
a transformation on the network output is applied.
}
\label{tab:xev_appendix}
\vspace{-.75em}
\end{table}

\section{Histograms of High-Level Features for CCD2 and CCD3}
\label{app:high_level_histograms}

\begin{figure}[t]
\centering
\includegraphics[width=0.98\textwidth]{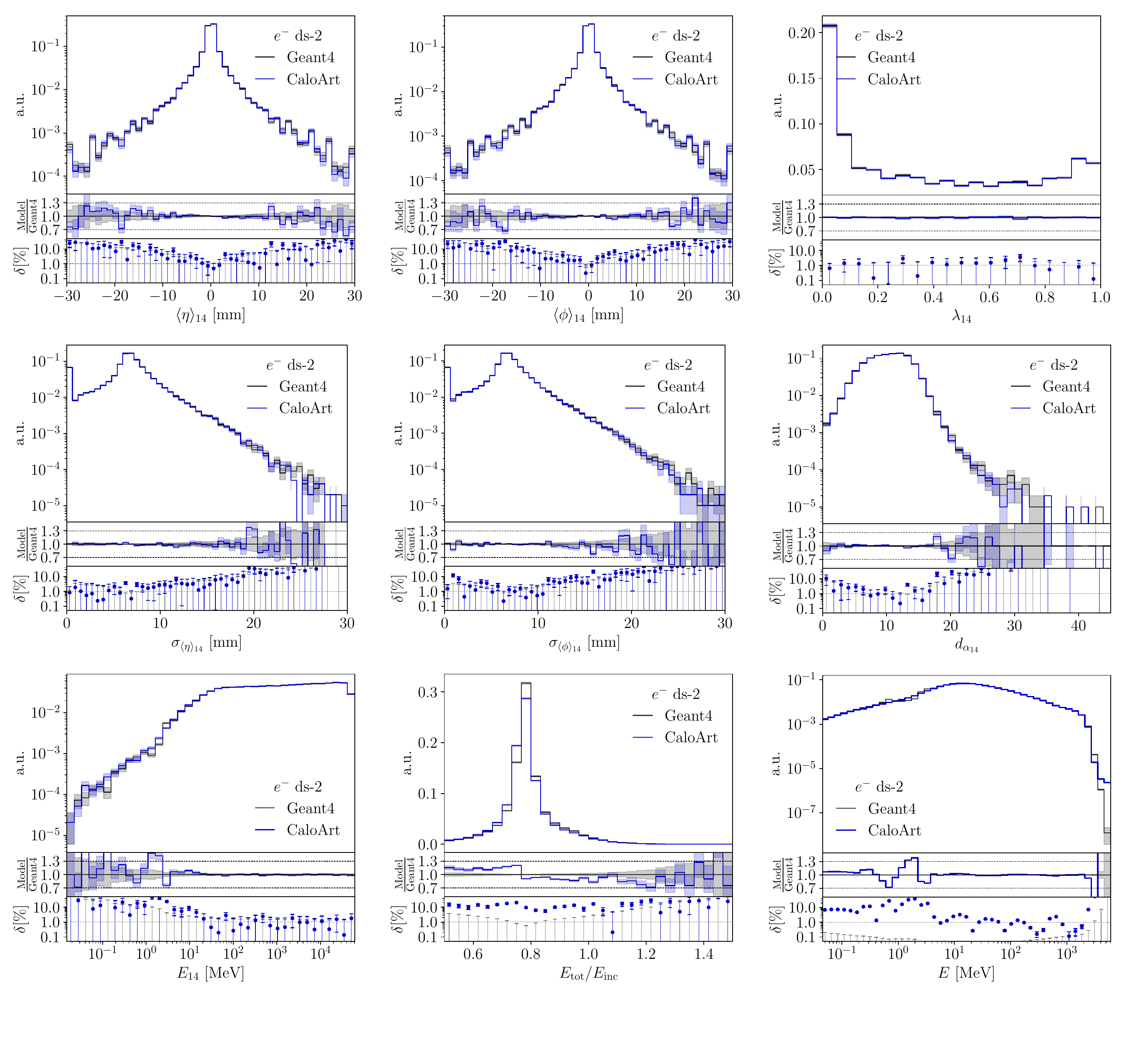}
\caption{
\textbf{Histograms of high-level features for the CCD2 detector.} From left to right and top to bottom: center of energy in the eta and phi directions, sparsity, width of the center of energy in the eta and phi directions, weighted depth by slice, layer energy depositions, total deposited energy ratio $E_{\mathrm{tot}}/E_{\mathrm{inc}}$, and voxel energy distribution.
}
\label{fig:ccd2_high_level_histograms}
\end{figure}

\begin{figure}[t]
\centering
\includegraphics[width=0.98\textwidth]{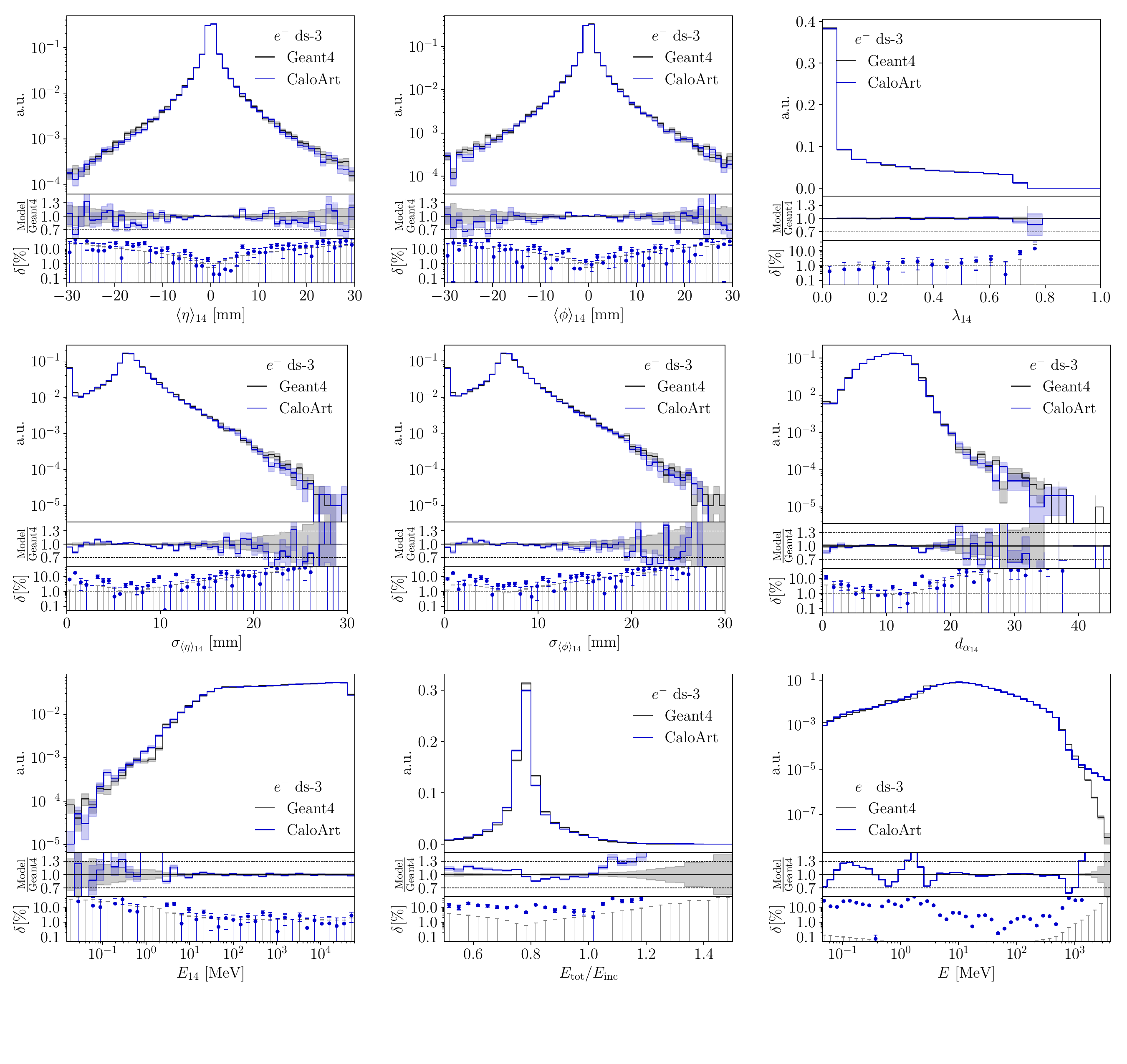}
\caption{
\textbf{Histograms of high-level features for the CCD3 detector.} From left to right and top to bottom: center of energy in the eta and phi directions, sparsity, width of the center of energy in the eta and phi directions, weighted depth by slice, layer energy depositions, total deposited energy ratio $E_{\mathrm{tot}}/E_{\mathrm{inc}}$, and voxel energy distribution.
}
\label{fig:ccd3_high_level_histograms}
\end{figure}

Figure~\ref{fig:ccd2_high_level_histograms} and Fig.~\ref{fig:ccd3_high_level_histograms} show the histogram comparisons of the high-level features for the final CCD2 and CCD3 models, respectively.

\clearpage

\section{Module-Level Validation of Gflops Counting}
\label{app:gflops_validation}

This appendix gives a compact module-level validation of the Gflops values
reported in the main text.
The original DiT paper reports Gflops, but does not spell out the exact
counting convention in the main text~\cite{2022_Peebles_arXiv_2212_09748}.
The convention is clarified in an official DiT GitHub discussion by one of the
DiT authors~\cite{2023_DiT_GitHub_Issue14}: most reported DiT Gflops values
refer only to the forward pass, and DiT follows the representation-learning
convention in which one multiply-add is counted as one flop.
We therefore report single-forward-pass Gflops under this DiT-compatible
convention throughout this paper.

The only convention conversion applied to the \texttt{DeepSpeed} profiling
output is a factor of two:
\begin{equation}
\mathrm{Gflops}_{\mathrm{DiT}} =
\frac{1}{2}\,\mathrm{Gflops}_{\mathrm{DS}},
\end{equation}
because \texttt{DeepSpeed} counts one multiply-add operation as two
floating-point operations. 
The module-level calculation below is performed under the same
DiT-compatible convention and provides a representative check of the reported
Gflops values.

\paragraph{Validation example.}
For CaloArt CCD3 configuration, the input volume is $(18,50,45)$ and the patch size is $(6,10,5)$, 
giving a token grid $(3,5,9)$ and sequence length $T=135$. 
The hidden dimension is $C=480$, the model has $L=6$ transformer blocks, $H=8$ attention heads, and head dimension $d_h=60$. 
The patch volume is $V_p=300$.

The dominant module-level MAC counts are:
\begin{align}
\mathrm{patch\ embed}
&= TCV_p
= 19.44 \times 10^6, \nonumber\\
\mathrm{attention/block}
&= T C(3C) + T C^2 + 2HT^2d_h
= 141.912 \times 10^6, \nonumber\\
\mathrm{SwiGLU/block}
&= T(480)(2560) + T(1280)(480)
= 248.832 \times 10^6, \nonumber\\
\mathrm{final\ projection}
&= TCV_p
= 19.44 \times 10^6.
\end{align}
Thus one transformer block contributes $390.744 \times 10^6$ MACs, and the six-block transformer stack contributes $2.344464 \times 10^9$ MACs. The smaller non-token-dominant terms are the timestep embedding, $0.35328 \times 10^6$ MACs, the three condition embedders, $0.08700 \times 10^6$ MACs, and the shared AdaLN modulation, $1.3824 \times 10^6$ MACs. This last term is small because the shared modulation is evaluated once per sample rather than once per token.

Summing these terms gives
\begin{align}
\mathrm{MACs}
=& \ 19.44 \times 10^6
+ 0.35328 \times 10^6
+ 0.08700 \times 10^6 \nonumber \\
&+ 1.3824 \times 10^6
+ 2.344464 \times 10^9
+ 19.44 \times 10^6 \nonumber \\
=& \ 2.38516668 \times 10^9.
\end{align}
Therefore, in the \texttt{DeepSpeed} convention,
\begin{equation}
\mathrm{Gflops}_{\mathrm{DS}}
= \frac{2 \times 2.38516668 \times 10^9}{10^9}
= 4.77033336.
\end{equation}
Equivalently, the DiT-compatible value used in the main tables is half of this number, $2.38516668$ Gflops. For the same forward pass, \texttt{DeepSpeed} reports $4.7714364$ Gflops in its convention. The difference from the module-level estimate is $0.00110304$ Gflops.
This is a relative difference of approximately $0.023\%$. The residual difference is attributable to small elementwise and functional operations not included in the main linear and attention MAC formulas, such as activation functions, normalization arithmetic, gating, and reshaping-related bookkeeping. The agreement validates the Gflops reporting protocol used in the main text.

\section{Classifier-Based Gflops Scaling on CCD2}
\label{app:ccd2_cls_Gflops}

For completeness, we also examine how the three classifier-based AUC metrics vary with backbone Gflops on CCD2. These plots complement the main-text FPD analysis in Fig.~\ref{fig:ccd2_fpd_Gflops_correlation} by showing the same compute-quality relation under classifier-based evaluation criteria.

\begin{figure}[t]
\centering
\includegraphics[width=0.32\textwidth]{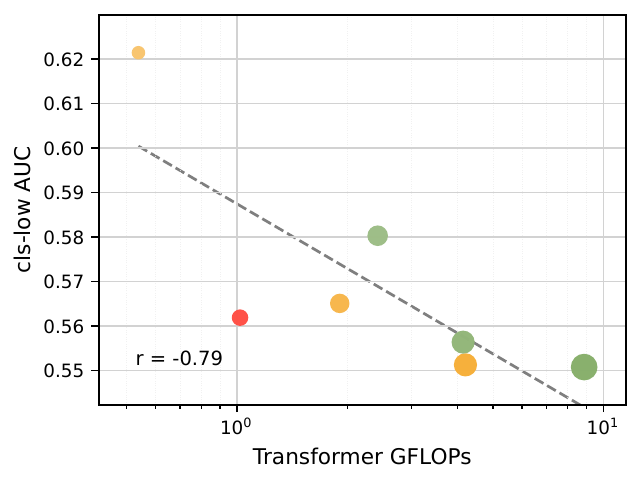}\hfill
\includegraphics[width=0.32\textwidth]{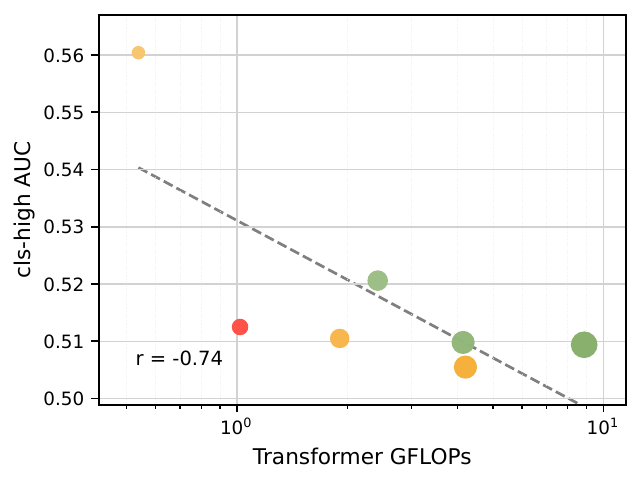}\hfill
\includegraphics[width=0.32\textwidth]{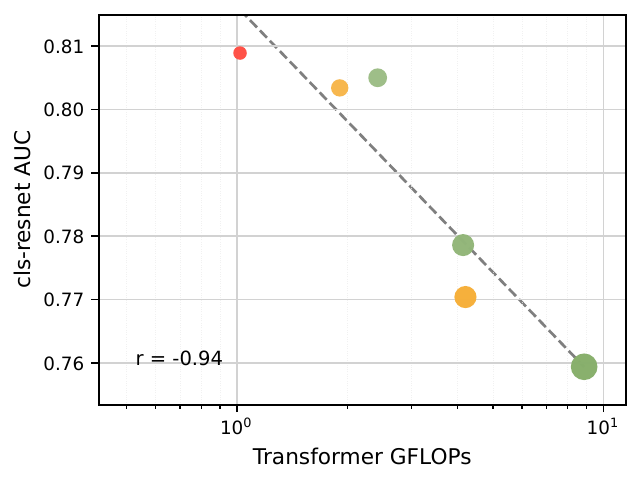}
\caption{
Classifier-based Gflops-scaling plots on CCD2. Left: cls-low AUC versus transformer Gflops. 
Middle: cls-high AUC versus transformer Gflops. 
Right: cls-resnet AUC versus transformer Gflops. 
These figures provide classifier-based counterparts to the FPD-versus-Gflops result shown in the main text. 
For the cls-resnet AUC panel, the visibly anomalous \texttt{h256-l6 | (3,8,5)} point is excluded from the fit-oriented visualization.
}
\label{fig:ccd2_cls_Gflops_correlation}
\end{figure}

\section{Published CCD3 Summary Used for the FPD--Time Pareto Plot}
\label{app:ccd3_public_submissions}

Table~\ref{tab:ccd3_public_summary} collects the published CCD3 entries used for the FPD--generation-time comparison in Fig.~\ref{fig:ccd3_fpd_time_pareto}. Geant4 is included for reference in the table, but is not plotted in Fig.~\ref{fig:ccd3_fpd_time_pareto} because the CaloChallenge source table does not report a GPU generation time~\cite{2025_Krause_RPP_88}.

\begin{table}[t]
\centering
\scriptsize
\setlength{\tabcolsep}{4pt}
\resizebox{\textwidth}{!}{%
\begin{tabular}{llccccc}
\toprule
& & \multicolumn{3}{c}{Classifier AUC} & FPD $\times 10^3$ & time [ms/shower] \\
\cmidrule(lr){3-5}
Model & Dataset & High-level & Low-level & ResNet & & \\
\midrule
Geant4 & ccd3 & $0.500 \pm 0.003$ & $0.498 \pm 0.002$ & $0.499 \pm 0.002$ & $8.8 \pm 0.6$ & -- \\
CaloDiffusion & ccd3 & $0.607 \pm 0.005$ & $0.561 \pm 0.003$ & $0.656 \pm 0.015$ & $71.2 \pm 1.9$ & $810.2 \pm 2.4$ \\
L2LFlows-MAF & ccd3 & $0.946 \pm 0.002$ & $0.720 \pm 0.016$ & $1.000 \pm 0.000$ & $665.5 \pm 1.7$ & $2454 \pm 120$ \\
conv.\ L2LFlows & ccd3 & $0.733 \pm 0.006$ & $0.588 \pm 0.004$ & $0.919 \pm 0.003$ & $171.6 \pm 1.8$ & $16.0 \pm 0.2$ \\
MDMA & ccd3 & $0.987 \pm 0.001$ & $0.944 \pm 0.002$ & $1.000 \pm 0.000$ & $588.6 \pm 2.5$ & $5.47 \pm 0.30$ \\
CaloClouds & ccd3 & $0.980 \pm 0.001$ & $0.865 \pm 0.005$ & $1.000 \pm 0.000$ & $948.2 \pm 4.6$ & $25.3 \pm 0.2$ \\
Calo-VQ & ccd3 & $0.998 \pm 0.000$ & $0.996 \pm 0.001$ & $1.000 \pm 0.000$ & $1193.9 \pm 2.8$ & $1.8 \pm 0.1$ \\
Calo-VQ(norm) & ccd3 & $0.994 \pm 0.000$ & $0.975 \pm 0.003$ & $1.000 \pm 0.000$ & $930.8 \pm 3.5$ & $26.6 \pm 0.2$ \\
CaloScore distilled & ccd3 & $0.924 \pm 0.002$ & $0.776 \pm 0.005$ & $0.994 \pm 0.001$ & $610.9 \pm 4.0$ & $162.2 \pm 0.5$ \\
CaloScore single-shot & ccd3 & $0.939 \pm 0.001$ & $0.807 \pm 0.005$ & $0.995 \pm 0.002$ & $584.0 \pm 2.9$ & $6.5 \pm 0.3$ \\
iCaloFlow teacher & ccd3 & $0.962 \pm 0.001$ & $0.911 \pm 0.003$ & $1.000 \pm 0.000$ & $897.6 \pm 5.3$ & $5596 \pm 56$ \\
iCaloFlow student & ccd3 & $0.971 \pm 0.001$ & $0.891 \pm 0.003$ & $1.000 \pm 0.000$ & $841.1 \pm 5.1$ & $16.7 \pm 0.5$ \\
Geant4-Transformer & ccd3 & $1.000 \pm 0.000$ & $0.886 \pm 0.011$ & $1.000 \pm 0.000$ & $22947.3 \pm 23.4$ & $8.77 \pm 0.36$ \\
CaloPointFlow & ccd3 & $0.945 \pm 0.002$ & $0.720 \pm 0.012$ & $1.000 \pm 0.000$ & $670.8 \pm 3.4$ & $5.52 \pm 0.03$ \\
CaloVAE+INN & ccd3 & $1.000 \pm 0.000$ & $0.881 \pm 0.005$ & $1.000 \pm 0.000$ & $11060.7 \pm 14.0$ & $3.83 \pm 0.09$ \\
CaloDREAM & ccd3 & $0.524 \pm 0.004$ & $0.630 \pm 0.005$ & $0.802 \pm 0.014$ & $20.7 \pm 1.1$ & $179.6 \pm 0.5$ \\
\bottomrule
\end{tabular}%
}
\caption{Summary of published CCD3 (ds3) submissions on common evaluation metrics. Classifier AUCs are taken from Table C23, FPD from Table C24, and timing reports GPU generation time per shower at batch size 100 from Table C30. Geant4 timing is not reported in the source table.}
\label{tab:ccd3_public_summary}
\end{table}

\clearpage
\addtocontents{toc}{\protect\setcounter{tocdepth}{2}}
\phantomsection
\addcontentsline{toc}{section}{References}
\bibliographystyle{JHEP}
\bibliography{main}

\end{document}